# Predictive Management of Electric Vehicles in a Community Microgrid

Bin Wang, Dai Wang, Rongxin Yin, Doug Black
Energy Storage and Distributed Resource Division
Lawrence Berkeley National Laboratory
Berkeley, CA, USA
{wangbin, daiwang, ryin, drblack}@lbl.gov

Cy Chan
Computational Research Division
Lawrence Berkeley National Laboratory
Berkeley, CA, USA
cychan@lbl.gov

***Abstract*—The charging load from Electric vehicles (EVs) is modeled as deferrable load, meaning that the power consumption can be shifted to different time windows to achieve various grid objectives. In local community scenarios, EVs are considered as controllable storage devices in a global optimization problem together with other microgrid components, such as the building load, renewable generations, and battery energy storage system, etc. However, the uncertainties in the driver behaviors have tremendous impact on the cost effectiveness of microgrid operations, which has not been fully explored in previous literature. In this paper, we propose a predictive EV management strategy in a community microgrid, and evaluate it using real-world datasets of system baseload, solar generation and EV charging behaviors. A two-stage operation model is established for cost-effective EV management, i.e. wholesale market participation in the first stage and load profile following in the second stage. Predictive control strategies, including receding horizon control, are adapted to solve the energy allocation problem in a decentralized fashion. The experimental results indicate the proposed approach can considerably reduce the total energy cost and decrease the ramping index of total system load up to 56.3%.**

## I. Introduction

The sales of plug-in electric vehicles (PEVs) continue to increase according to the latest report [1], where pure electric vehicles (EVs) account for 54.3% of the total PEV sales across United States in 2016. The challenges of integrating EVs into the local distribution grid, with the increasing level of penetration, have been identified and studied by previous research work [2]–[4]. For instance, unregulated EV charging behaviors may cause degradation of the power quality in the local distribution grid [2], and increase the total energy bill for EV drivers, and the local system operators/aggregators [4], who purchase energy from the market and provide charging services to numerous EVs in the local community. Other than pure theoretical research, more challenges concerning real-world implementations are observed recently, such as the shape of the baseload profile in the local community, uncertainties of driver behaviors [5], [6], etc.

Charging load from Electric vehicles (EVs) is defined as deferrable load [7], which can be controlled and shifted to a different time windows in order to serve a myriad of grid objectives, such as valley-filling, load following [8] and cost minimization [5], [6], [9]. Local renewable generation together with time-of-use (TOU) prices are considered in [5], [6], but the scale of the simulated system is not sufficiently large to participate the wholesale energy market, thus day-ahead operations cannot be performed to further improve the cost performance. Authors in [10] utilize model predictive control to handle the uncertainties of aggregated EV energy demand together with building load. However, the flexibilities of individual EVs are not formulated, thus, the methodology cannot be directly adapted to real-world cases. Instead of aggregated control of EVs, [9] discusses an implementable price-based approach, allowing EV drivers to select a preferred energy price threshold, based on which energy allocation and sharing strategies are determined. In addition, EVs, BESS, building loads and physical constraints of the local distribution grid, are considered collaboratively in the microgrid operations in order to minimize the total operational cost [11], [12].

In this paper, we formulate a two-stage EV charging control problem, i.e. 1) energy purchase from day-ahead energy market using the forecast baseload, solar generation and EV load profile; 2) real-time predictive energy allocations considering uncertainties of stay duration and energy consumption for each individual EV. Compared with previous research, the contributions of this paper are threefold: 1) Profiles of community netload and renewable generation in a real-world local community, i.e. campus of Cornell University [13], are utilized to setup the experiment; 2) The samples of EV charging sessions in the experiment are randomly drawn from the distribution of charging behaviors in a real-world EV implementation project, which increases the fidelity of our simulations; 3) Two-stage operations, i.e. day-ahead operations (energy transaction from day-ahead market) and real-time predictive allocation, are combined to regulate the EV charging behaviors, considering the uncertainties within individual EV charging sessions. The real-time strategy allocates the energy to each individual EV in a decentralized fashion. The proposed approach leads to a considerable reduction of total operational cost and more than 50% reduction of system ramping index;

The paper is organized as follows: In section II, we show the system architecture and overview of the time scales in the two-stage scheduling strategy; Problem formulation and component modeling are provided in section III; Section IV presents the experiment results as well as the related discussions. Finally, we conclude the paper in section V with potential future work.

## II. Proposed System Architecture

In the proposed system architecture, a community microgrid is assumed to be connected with advanced metering

This work was funded by California Energy Commission (CEC) Contract 14-057, and LDRD funding at Computational Research Division (CRD), Lawrence Berkeley National Laboratory.

infrastructure (AMI), which enables the exchange of the timely measurement data and control signals between the controller and the various microgrid components. To study the performance of the proposed approach with predictive management strategies over EVs, other types of microgrid components, such as distributed generators, battery energy storage system (BESS) and other devices, are captured in the baseload rather than individually modeled, as the focus is on the behavior of the EV controller. Under this architecture, the two-stage operation is proposed, i.e. day-ahead operation, which produces day-ahead energy consumption schedules given daily estimated values of load and generation, and predictive real-time energy allocation program, that runs in smaller time scales and delivers the charging energy to individual EVs, considering the uncertainties of EV travel itineraries and energy demand. During allocation stage, system state estimation, optimization and implementation of charging plans are performed consecutively in each time step until the end of the scheduling horizon. In addition, we assume that all microgrid components are under the same distribution feeder, which simplifies the load management.

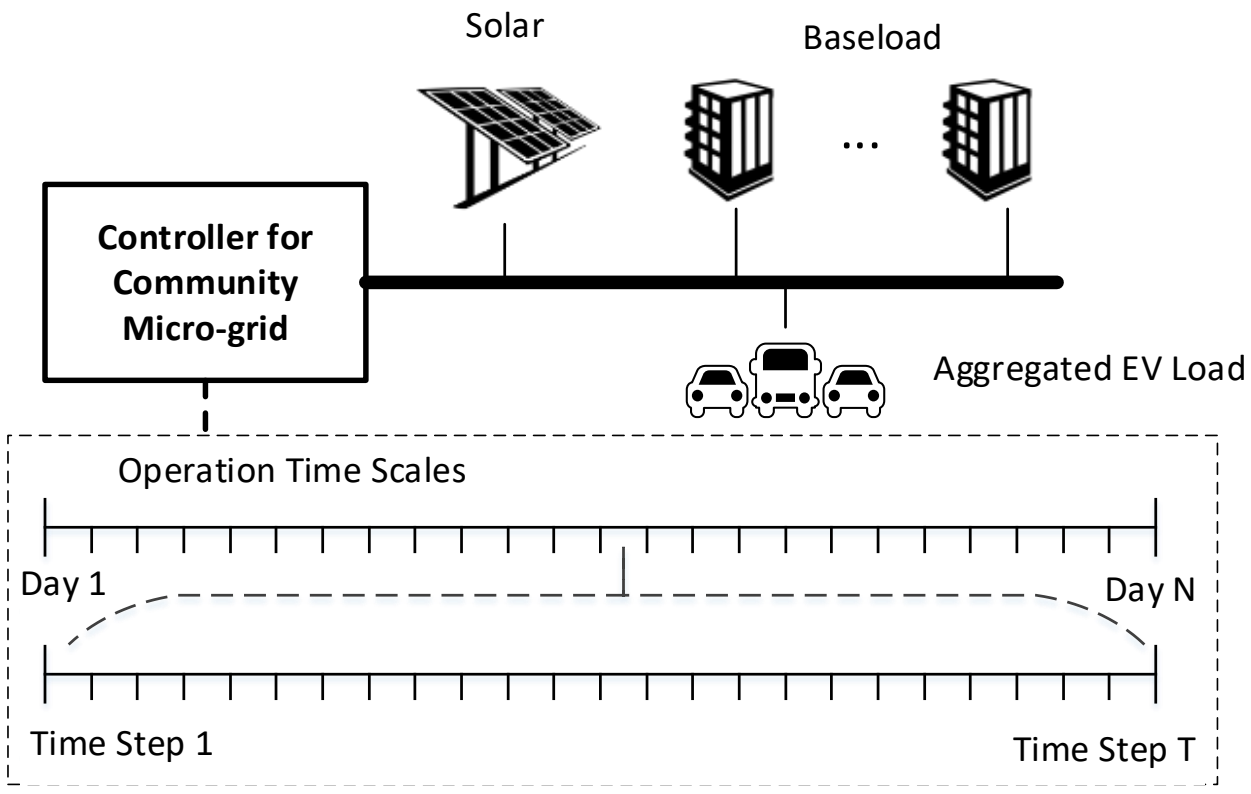

Figure 1 System architecture

## III. SYSTEM MODELING

### A. Aggregate level optimization

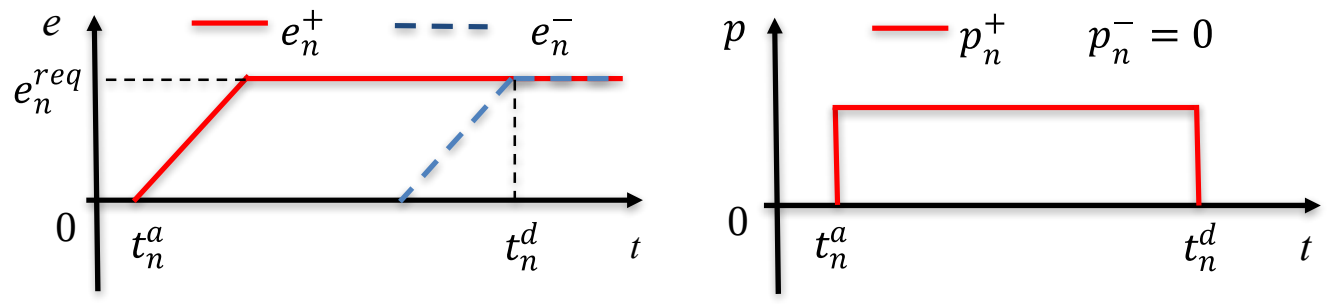

(a) Energy Boundaries (b) Power Boundaries

Figure 2 Energy and power boundaries of an individual PEV.

Before quantifying the aggregated flexibility of EVs in the community microgrid, we use the methods in [14] to model the power and energy boundaries of individual EV, which has critical impact on the EV integration into market. The charging flexibility of a given PEV $n$ can be modeled by its energy and power boundaries, i.e. { $e_n^{+/(-)}$, $p_n^{+/(-)}$}, as shown in Figure 2. These boundaries form enclosed regions, which specify the set of all feasible charging trajectories. $t_n^s$ is the start/arrival time of PEV $n$. $t_n^d$ is the departure time of PEV $n$. $e_n^{req}$ denotes the energy demand by PEV $n$. The upper energy boundary $e_n^+$ represents the fastest path, i.e. as soon as possible, to charge the battery, whereas the lower boundary $e_n^-$ represents the slowest path, i.e. as late as possible, to charge the battery. $p_n^+$ is the maximum charging power that the charging facility can provide. If we do not consider energy discharging, the power lower boundary is always 0.

For a population of PEVs, we add energy and power boundaries of individual EV together to define the aggregate flexibility in equation (1) – (4):

$$E^{+/(-)}(t) = \sum_i e_i^{+/(-)}(t), \qquad \forall t \in [0,T] \tag{1}$$

$$P^{+/(-)}(t) = \sum_i p_i^{+/(-)}(t), \qquad \forall t \in [0,T] \tag{2}$$

$$P^-(t) \le P(t) \le P^+(t), \forall t \in [0,T] \tag{3}$$

$$E^-(t) \le \sum_{\tau=0}^{t} P(\tau)\cdot \Delta t \le E^+(t), \forall t \in [0,T] \tag{4}$$

where $E^+(t)$ and $E^-(t)$ are the aggregated energy boundaries. $P^+(t)$ and $P^-(t)$ are the power boundaries. These power and energy boundaries will work as important constraints in the aggregator's optimization problem. With the above modeling, we change the large-scale, discrete, randomly distributed individual charging demands into a single, smooth and comparatively steady storage-like aggregate model.

In a community, solar energy is assumed at zero marginal cost and is used in priority. Since the solar energy cannot cover total energy demand, the remaining load is supplied by the main distribution grid. In this paper, we assume that the community is exposed to the wholesale electricity market, where transactions can be made in the day-ahead energy market according to the forecast information on its base load, solar generation and PEV charging load. The aggregator aims at minimizing its total energy cost. Meanwhile, the aggregator also wants to smooth the internal net load to reduce the impact on the main grid and reduce the impact of the load ramping. We formulate first-stage optimization problem for the aggregator as:

$$\min \sum_{t=1}^{T-1} \lambda(t)\cdot\left[\hat{B}(t) - \hat{S}(t) + P(t)\right] + \theta \cdot \sum_{t=1}^{T-1}\left[\left(\hat{B}(t+1) - \hat{S}(t+1) + P(t+1)\right) - \left(\hat{B}(t) - \hat{S}(t) + P(t)\right)\right]^2 \tag{5}$$

subject to (3) and (4),

where $t$ is the time index, $t = 1,\ldots,T$; $\lambda(t)$ is the wholesale energy price at time $t$; $\hat{B}(t)$ is the forecast base load at time $t$; $\hat{S}(t)$ is the forecasting solar generation at time t; $P(t)$ is the aggregate charging power at time $t$; $\theta$ is the weigh factor to adjust the aggregator's expectation on ramping mitigation. In the objective function, the first term is the total energy cost, while the second term represents the maximum ramping rate of the netload. Equation (3) is the power constraint, and equation (4) is the energy constraint to guarantee all the EVs have enough energy for their driving demand.

## B. Decentralized energy allocation to individial vehicles

During the day-ahead operations, the optimal EV charging schedule is determined by solving (5). The task for the second-stage operation is to follow/track the day-ahead optimal charging load profile. Allocating energy to each individual EV requires the controller to know the exact session parameters, including $t_n^s$, $d_n$ and $e_n$. In addition, enough energy is supposed to be delivered to each individual EV before its deadline $t_n^l$, i.e. the leave time of the EV $n$. However, in the real-world case, the charging session parameters are not fixed, i.e. leave time, i.e. $t_s + d$, and energy demand $e_n$ are not certain. For each EV, the charging power and battery energy constraints are modified with itinerary uncertainties as follows:

$$0 \le p_n(t) \le p^+ \cdot \eta, \qquad \forall t \in [t_n^s, t_n^s + \hat{d}_n] \tag{6}$$

$$e_n(t) = e_n(t - \Delta t) + p_n(t) \cdot \Delta t, \ \ \forall t \in [t_n^s, t_n^s + \hat{d}_n] \tag{7}$$

$$e_B \ge e_n\left(t_n^s + \hat{d}_n\right) \ge \hat{e}_n \tag{8}$$

where $p_n(t)$ is the charging power of EV $n$ at time $t$; $p^+$ denotes the maximum charging rate of the charging infrastructure and $\eta$ is the power efficiency. Note that, the leave time of EV $n$ is replaced with $t_n^s + \hat{d}_n$, where $\hat{d}_n$ is the estimated stay duration from an online predictor. Equation (7) describes the increment of stored energy in the battery of each individual EV. By the end of the scheduling horizon, the stopping criteria for each EV, i.e. equation (8), is to get more energy than the estimated value $\hat{e}_n$, while below the total battery capacity $e_B$.

Given the optimal EV charging load profile by the aggregator level optimization, the real-time allocation aims to minimize the summed difference between the desired and real netload curves. Thus, the objective of the second-stage predictive energy allocation problem is formulated as follows:

$$\min \sum_{t=\tau}^{T-1} \left[ \left(\hat{B}(\tau) - \hat{S}(\tau) + \hat{P}(\tau)\right) - (\hat{B}'(\tau) - \hat{S}'(\tau) + \sum_{n=1}^{N} p_n(\tau)) \right]^2 \tag{9}$$

subject to: (6), (7) and (8).

As the number of EVs increases, it is required by the centralized controller to: i) collect the timely measurements for all EVs, ii) compute the optimal charging schedules and then iii) send control signals to each individual charging facility. Compared with the decentralized counterparts, it has apparent drawbacks, including 1) user privacy issue, i.e. schedule related information for all users will be collected altogether by the centralized server; 2) high computation burden and communication network condition may delay the delivering of the optimal charging schedules. Instead in decentralized approaches, each individual EV computes its own optimal charging schedules, given only the control/price signals from a central server, without knowing schedule information of other EVs. EVs send their updated charging schedules to the server asynchronously and get updated control/price signals back until all EVs reach an equilibrium state. We adapt the decentralized algorithms developed in [8], [15], and integrate the real-time energy allocation with day-ahead operations. We extend the scheduling algorithm to follow day-ahead EV charging load profile and add another layer of iterations to simulate the real-time operations, i.e. predictively optimize the energy consumption schedules in each time step given estimated session parameters for each EV, and then implement only the first element in the charging schedule (the next time step). The scheduling service will continue until the end of the time horizon. The modified decentralized algorithm is as follows:

*a) For aggregator:* In each iteration $k$, the aggregator or system operator needs to calculate the consensus-based control signal, i.e. $c_k^\tau$, as equation (10), assuming there's updated the forecast of $B(\tau)$ and $S(\tau)$, i.e. $\hat{B}'(\tau)$ and $\hat{S}'(\tau)$. As the focus is on the EV behaviors, we assume the perfect forecast of baseload and solar:

$$c_k^\tau = \frac{1}{\beta \cdot N} \cdot (\sum_{n=1}^{N} p_k^\tau - \hat{P}(\tau)) \tag{10}$$

where $\hat{P}(\tau)$ is the optimal aggregated charging load at time $\tau$ from the first-stage planning and $\beta$ is a Lipschitz constant.

*b) For each EV:* Each EV solves the following local optimization problem given its estimated itinerary and the energy demand values, as well as the updated control signal $c_k^\tau$ from the system operator.

$$\min \sum_{t=\tau}^{T} c_k^\tau \cdot p_n(t) + \sum_{t=\tau}^{T} [p_n^{k-1}(\tau) - p_n^k(\tau)]^2 \tag{11}$$

subject to: (6), (7) and (8).

The detailed approach is illustrated as follows:

| **Predictive & Decentralized EV Management Strategy** |
|---|
| **Day-ahead operation:** |
| Solve equation (5), subject to (3) – (4); |
| **Real-time operation:** |
| **For** $\tau = 1$ $to$ $T$ |
| Retrieve forecast baseload $\hat{B}(\tau), \hat{B}(\tau+1), \cdots, \hat{B}(T)$ and |
| solar data $\hat{S}(\tau), \hat{S}(\tau+1), \cdots, \hat{S}(T)$; |
| **Do** |
| Initialize a random charging schedule for each EV, i.e. |
| $p_n^0(t_n^s), p_n^0(t_n^s+1), \cdots, p_n^0\left(t_n^s + \hat{d}_n\right)$; |
| Iteration count $k = 0$; |
| Operator: calculate $c_k^\tau$, using (10); |
| **For** $n = 1{:}N$ |
| Estimate the updated stay duration $\hat{d}_n$ and energy |
| demand $\hat{e}_n$; |
| Each EV: solve (11) for updated schedule, i.e. |
| $p_n^k(t_n^s), p_n^k(t_n^s+1), \cdots, p_n^k\left(t_n^s + \hat{d}_n\right)$, subject to (6) - (8); |
| **End For** |
| $k = k + 1$; |
| $error = \lVert c_k^\tau - c_{k-1}^\tau \rVert$; |
| **While** $error > \underline{err}$ |
| **For** $n = 1{:}N$ |
| Implement $p_n(\tau)$, if $t_n^s \le \tau <$ if $t_n^s + \hat{d}_n$ |
| **End for** |
| **End for** |

Note that, the stay duration and energy consumption values may vary as the time step moves forward in the simulation, since each EV driver does not have to stick to a fixed charging plan. In the real-time allocation stage, the controller for each EV will update the estimation of these sessions parameters, i.e. line 13-14 in the above algorithm, using approaches such as the method based on K-Nearest-Neighbors in [16] and kernel-based one in [5]. For each time step, the consensus based algorithm, i.e. from step 7 – 20, will determine the equilibrium control signals that achieves global optimal, while the itinerary and energy demand constraints of each EV can also be satisfied.

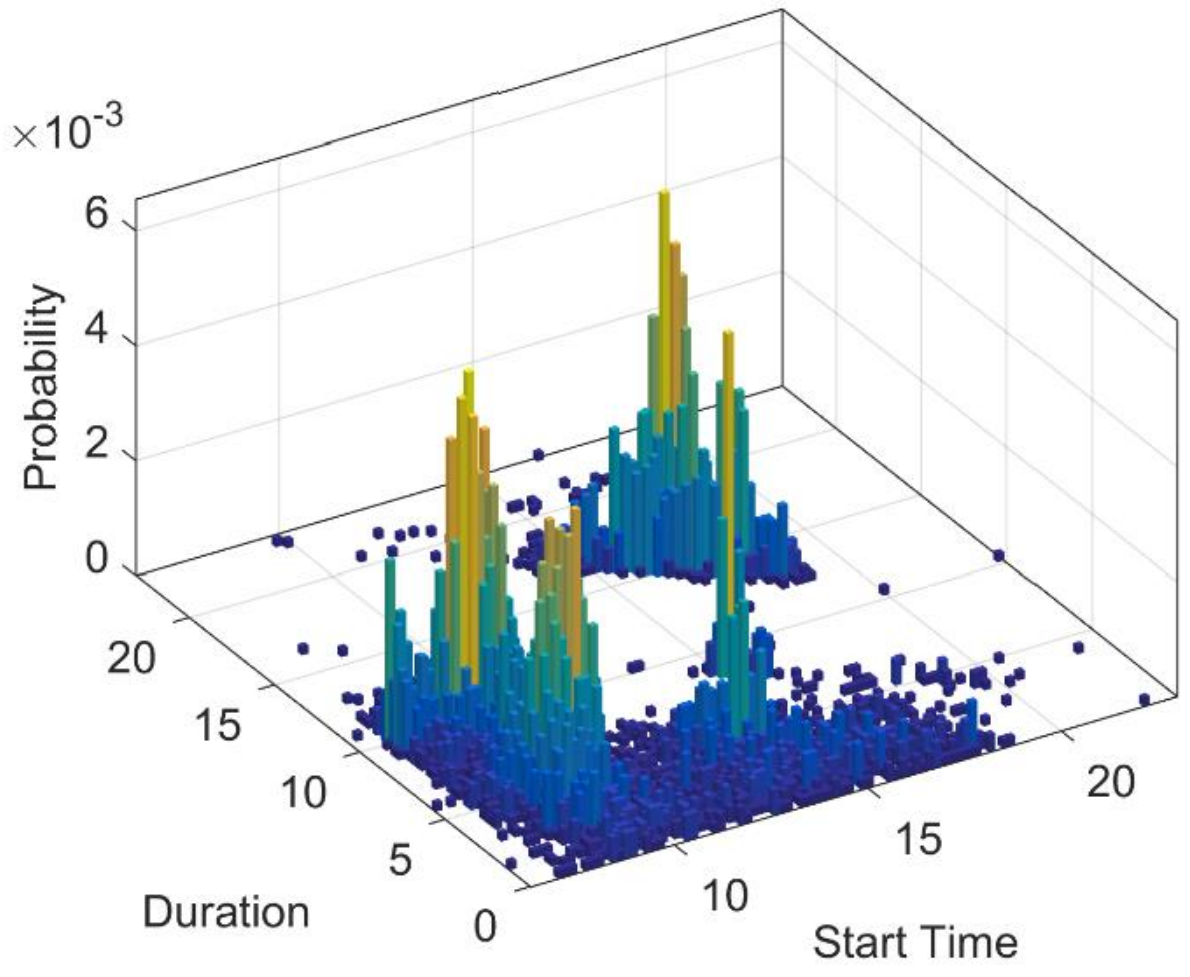

Figure 3 Distribution of Charging Behaviors

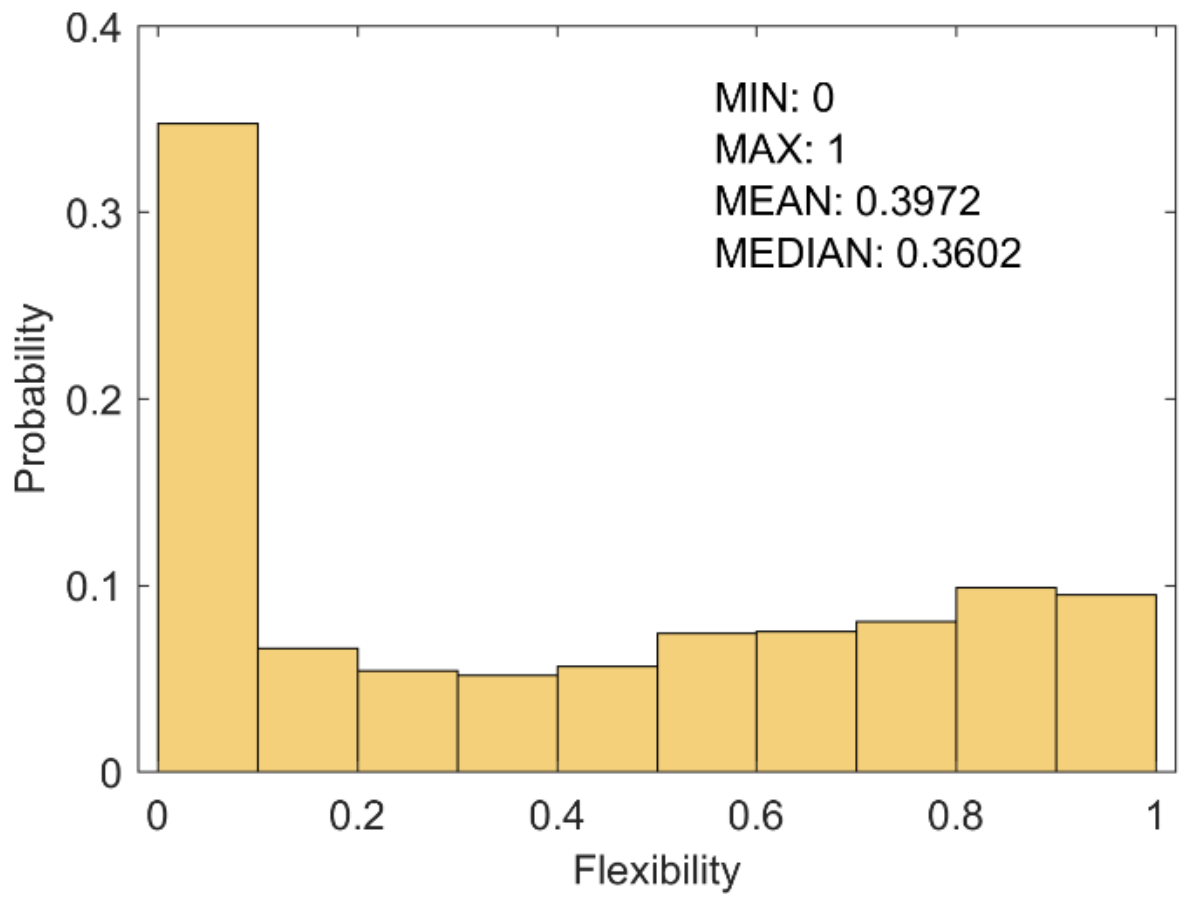

Figure 4 Flexibility of Charging Sessions

## IV. Experiment Results and discussion

### A. Experiemnt Setup

The datasets containing EV charging behaviors are from the smart charging demonstration project located in the county of the Alameda, north California. Charging sessions over 1 year period are extracted from database to generate the distributions of driver behaviors. To represent each charging session, three critical parameters, i.e. start time, stay duration (plugged) and the energy demand, are generated based on the distribution shown in the Figure 3 and Figure 4. We use session flexibility, which is defined as the ratio of charging time within the total plugged-in time, to generate the energy demand values. As illustrated in Figure 4, the higher the session flexibility, the higher the degree of freedom we can defer or optimize charging load. To further simulate variability across EVs, perturbations are added to the stay duration and the energy demand value for each EV.

A sample of the community microgrid netload and solar photovoltaic (PV) generations from the campus of Cornell University is utilized in this paper to verify the performance of the proposed scheduling system. Random perturbations are added to the netload and PV generation curves to represent the real-time estimated values. In addition, sample price data from CAISO wholesale energy market is used in the simulations.

### B. Results and Discussion

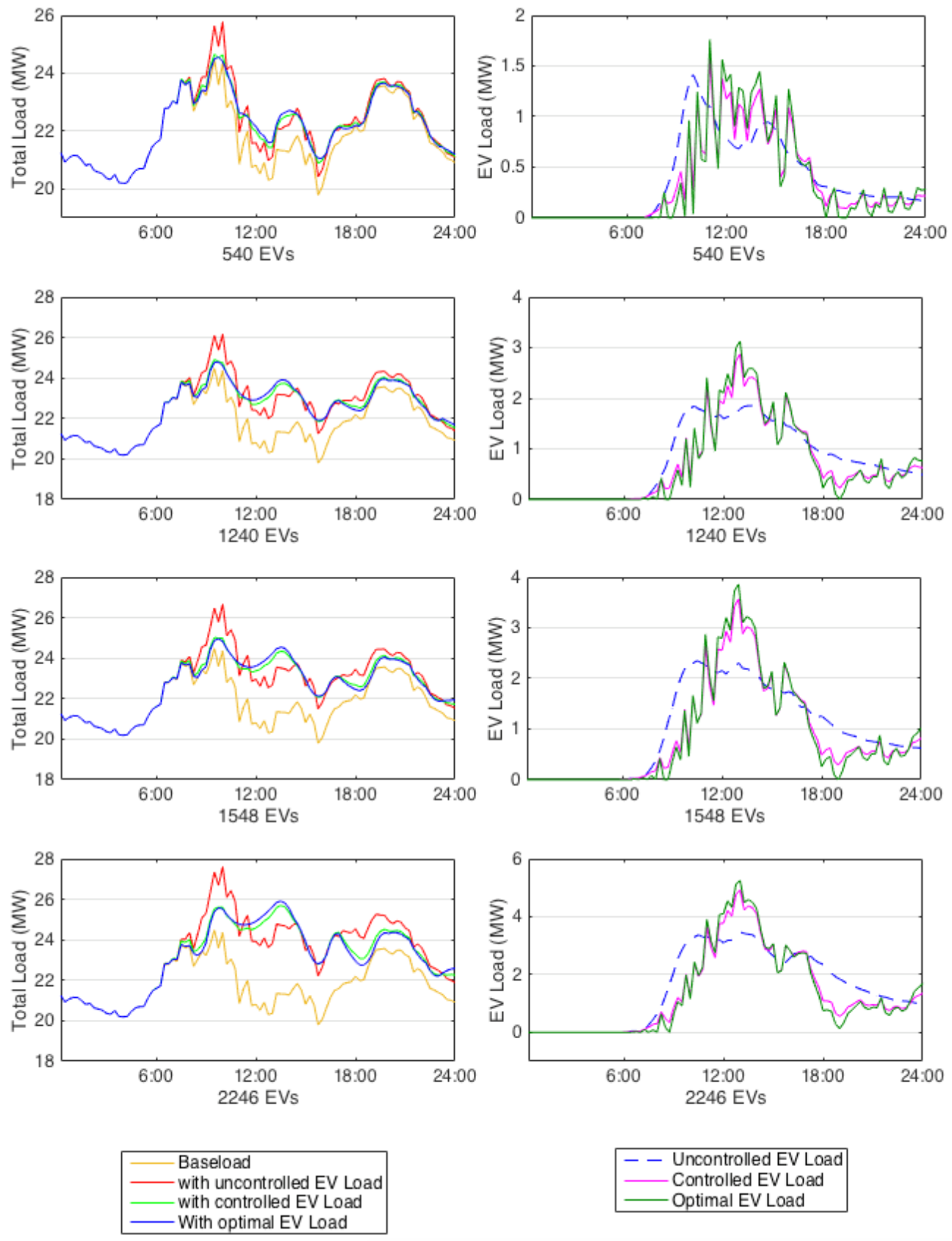

Figure 5 Scheduling results with different number of EVs in each case

Using the distribution of the data collected from the real-world EV implementation projects, we generate three experimental scenarios with different number of EVs (penetration), i.e. 540, 1240, 1548 and 2246. Each EV has a varied set of session parameters, including start time, stay duration and energy demand. In each scenario, the scheduling results are shown in Figure 5, where uncontrolled EV load, controlled EV load, community baseload, baseload with EV load are displayed. The blue solid lines in Figure 5 denote total system load, with the optimal day-ahead EV load on top of the baseload, while the green ones represent the real-time total system load with EV load controlled by the proposed predictive scheduling strategies. From the charts, one can find that the green curves follow and track the blue ones very well, even with small mismatches caused by the randomized driver behaviors. The performance of load following scales well as the number of simulated EVs increases. With optimization, the total system load peak can avoid the time windows around 10 AM, reducing additional operational cost.

To evaluate how effectively the proposed strategies can shift the EV load into the time window with lower wholesale energy prices, we compare the energy purchase cost in different scenarios in Figure 6. The blue circles denote the total energy cost from wholesale market with the optimal EV charging load, and the red stars indicate the cost by the proposed predictive scheduling strategies. One can find that the controlled EV load follows the optimal EV load profile, and contributes to the reduction of the total energy cost.

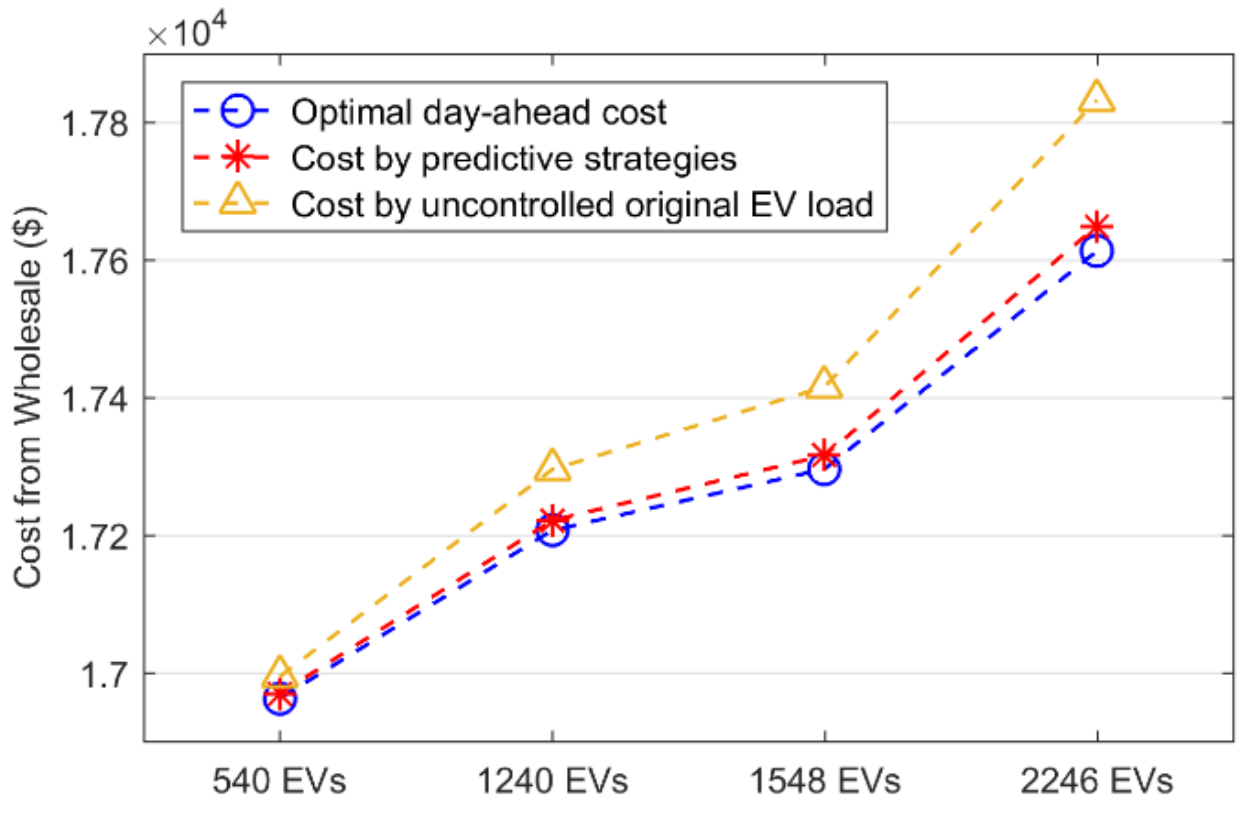

Figure 6 Total system cost from wholesale market

In the objective of the day-ahead operation, i.e. equation (5), we also take into account the ramping of the community netload. In other words, the load fluctuation between consecutive time steps should be reduced by tuning the coefficient $\theta$. The system load ramping statistics are shown in Table 1. In the case with 1240 EVs, the system netload with controlled EV load profile has a decreased maximum load ramping index from 1800 to 786.2, which is 56.3% reduction. Thus, the capability of the proposed approach in reducing the netload ramping was demonstrated.

Table 1 System Load Ramp Reduction ($\theta = 0.01$)

| **Ramp Reduction** | **540** | **1240** | **1548** | **2246** |
|---|---|---|---|---|
| ***With Uncontrolled EV Load*** | 1768.3 | 1800.0 | 1781.8 | 1737.4 |
| ***With Controlled EV load*** | 941.4 | 786.2 | 793.2 | 889.7 |
| ***Max. ramp reduction*** | 46.8% | 56.3% | 55.5% | 48.8% |

## V. Conclusion and Future Work

In this paper, we developed a predictive management strategy that combines day-ahead operations in the wholesale energy market and real-time energy allocations to community EVs. The experimental results demonstrated the effectiveness of the proposed strategy in reducing the netload load ramping and the total energy cost from wholesale market. For future work, retail market integration and power system topology will be considered to further refine the scheduling strategies.

## Acknowledgment

The research that leads to the results in this paper has been supported by California Energy Commission (CEC) Contract 14-057. Lawrence Berkeley National Laboratory is operated for the US Department of Energy under Contract Grant No. DE-AC02-05CH11231. In addition, the work is partially supported by ongoing LDRD funding at Computational Research Division (CRD) for ExaGrid project, Lawrence Berkeley National Lab.